\documentclass[doublecol]{epl2} 
\title{Correlational Origin of the Roton Minimum}
\shorttitle{Roton Minimum} 
\author{
  G. J. Kalman\inst{1} \and
  P. Hartmann\inst{1,2} \and
  K. I. Golden\inst{3} \and
  A. Filinov\inst{4} \and
  Z. Donk\'o\inst{1,2}
}
\shortauthor{G. J. Kalman \etal}

\institute{                    
  \inst{1} Department of Physics, Boston College - Chestnut Hill, 
  MA 02467, USA \\
  \inst{2} Research Institute for Solid State Physics and Optics 
  of the Hungarian Academy of Sciences - H-1525 Budapest, POB 49. 
  Hungary \\
  \inst{3} Department of Mathematics and Statistics, Department of 
  Physics University of Vermont, Burlington, Vermont 05401-1455, USA \\
  \inst{4} Institut f\"ur Theoretische Physik und Astrophysik, Christian-
Albrechts-Universit\"at zu Kiel, D-24098 Kiel, Germany  
}

\abstract{ We present compelling evidence supporting the conjecture that the 
origin of the roton in Bose-condensed systems arises from strong correlations 
between the constituent particles. By studying the two dimensional bosonic dipole 
systems a paradigm, we find that classical molecular dynamics (MD) simulations 
provide a faithful representation of the dispersion relation for a low-
temperature quantum system. The MD simulations allow one to examine the effect of 
coupling strength on the formation of the roton minimum and to demonstrate that 
it is always generated at a sufficiently high enough coupling. Moreover, the 
classical images of the roton-roton, roton-maxon, etc. states also appear in the 
MD simulation spectra as a consequence of the strong coupling. }

\begin{document}

\maketitle

Since the appearance of the pioneering papers by Landau on the theory of 
superfluidity \cite{1}, the notion of the "roton minimum" in the collective mode 
dispersion of the system has played a pivotal role in the explanation of the 
behavior of the superfluid phase of $^4$He. This Letter addresses this old issue 
from a novel point of view, made possible by results from newly available 
computer simulation techniques.

The revealing feature of the superfluid behavior is the non-viscous flow of the 
liquid \cite{2} (or the frictionless motion of an impurity in the liquid \cite
{3}) below the critical temperature $T_c = 2.17$~K, and the breakdown of this 
behavior above a critical flow velocity. In Landau's theory, it is postulated 
that in order to explain this breakdown, the $\omega(k)$ dispersion curve of the 
collective	excitation	must	 be non-monotonic, starting with a longitudinal $
\omega = ks$ acoustic portion, reaching a maximum (the "maxon"), which is then 
followed by a deep minimum around a $k = k_{\rm min}$ (the "roton minimum"), such 
that the $\omega = ks_{\rm max}$ line, representing the relative motion with 
respect to the fluid, becomes tangent to the dispersion curve in the neighborhood 
of the minimum.

Since 1947 the physical origin of the roton minimum and the question whether it s 
a part of thelongitudinal  phonon eexcitation has been open to conflicting 
interpretations. Landau's original suggestion was that the development of the 
roton minimum had to be sought in the existence of a new collective excitation, 
quite distinct in nature from the longitudinal acoustic phonon and rather due to 
a rotation-like collective motion of the fluid: hence, of course, the term by 
which this phenomenon has become to be known.

Almost a decade later Feynman and Cohen \cite{4} took up the detailed analysis of 
the ground state of a strongly correlated bose liquid. Their seminal work led to 
the identification of $S(k)$, the static	 structure function,	 as the centrally 
important quantity in the characterization of the system. Through a series of 
physical arguments - whose outcome today would be summarized as the "Feynman 
Ansatz" - Feynman was able to relate the ground state $S(k)$ to the ground state 
excitation spectrum $\omega(k)$ of the fluid through his famous formula
\begin{equation}
\label{eq.1}
\omega(k)=\frac{\hbar k^2}{2 m S(k)}.
\end{equation}
Based on (\ref{eq.1}), the observation that strong correlations create a sharp 
maximum in $S(k)$ resulted in the re-emergence of the roton minimum from a rather 
different physical foundation than that envisioned by Landau. Nevertheless, 
perhaps surprisingly, the interpretation offered by Feynman still endorsed the 
association of the roton minimum with a vortex-like excitation.

In the subsequent decades while observational data through neutron scattering 
measurements accumulated and unambiguously revealed the existence of the roton 
minimum \cite{5} at around $k_{\rm min} = 1.93$~{\AA}$^{-1}$ almost as predicted 
by Landau's hypothesis, the interpretation of its physical origin remained 
somewhat in a limbo. The idea that the low-$k$ phonon and the high-$k$ roton stem 
from different sources prevailed: in a scenario developed by Glyde and Griffin in 
1990 the roton was viewed as a single particle excitation mixed with the density 
fluctuation spectrum \cite{6}. At the same time, however, in the 1970-s Schneider 
and co-workers \cite{7} pointed at a different role that the roton minimum seemed 
to play: namely being the heralded of an incipient liquid-solid phase transition. 
Still, it took an additional thirty years for the corollary of this statement to 
be enunciated by Nozieres \cite{8}, by suggesting that the physical mechanism 
responsible for the very creation of the roton minimum is the strong correlations 
prevailing in the system (although the fact that local minima appear in the 
dispersion curves of other strongly correlated systems resembling the $^4$He 
roton minimum was realized much before that \cite{9}). The purpose of this Letter 
is to provide evidence, based on recent computer simulations, that favors this 
latter physical picture.

New developments in condensed matter physics over the past decade, notably the 
emergence of experimentally available novel 3D and 2D bosonic systems, have 
created the feasibility of studying the physics of the roton minimum under more 
controlled conditions. Perhaps the most interesting new system that has appeared 
along these lines is the two-dimensional liquid of excitonic dipoles. The 
existence of an excitonic liquid in 2D semiconductor layers and its expected 
condensation at $T=0$ was predicted by Keldysh and co-workers a long time ago 
\cite{10}. Subsequently, the possibility of realization of stable excitons in 
electron-hole bilayers was analyzed by Lozovik and others \cite{11}. It is now 
fairly well established that with a sufficiently small layer separation $d$ ($d/
a<1$, where $a$ is the average interparticle distance within a layer) the 
description of these excitons by a model of a single 2D layer of point dipoles 
interacting through a classical dipole-dipole potential is a very good 
approximation.

Recently, the properties of such a 2D system of bosonic dipoles, both at zero and 
at finite temperatures have been the subject of intensive analytic and computer 
simulation efforts. Some experimental observations in GaAs semiconductor bilayers 
\cite{12} can also be interpreted in terms of this model. Contrasting with liquid 
$^4$He, the unique feature of the dipole system resides in the simplicity of the 
interparticle interaction and the possibility of tuning the coupling strength by 
changing system parameters, such as temperature and density. It is now feasible 
to trace both the equilibrium behavior and the structure of collective 
excitations over a wide range coupling strengths and temperatures, by applying 
various techniques of computer simulation \cite{13,14}. This includes the 
analysis of the generation of the roton minimum and of its evolution as a 
function of the system parameters. These simulation studies have shed a new light 
on this old issue. In the following we discuss what can be gleaned from these 
recent results and point out what conclusion can be drawn from them as to the 
physical origin of the roton minimum.

The 2D point-dipole system can be described as a collection of $N$ spinless point 
dipoles, each of mass $m = m_{\rm e} + m_{\rm h}$, occupying the large but 
bounded area $A$; $n = N/A$ is the average density. The dipoles are free to move 
in the $x-y$ plane with dipolar moment oriented in the $z$ direction. The 
interaction potential is accordingly given by
\begin{equation}
\label{eq.2}
\varphi(r)=\mu^2/r^3,
\end{equation}
where $\mu$ is the electric dipole strength. The coupling strength can be 
characterized at arbitrary temperature by
\begin{equation}
\label{eq.3}
\tilde\Gamma_D =\mu^2/a^3E_{\rm kin},
\end{equation}
where $E_{\rm kin}$ is a characteristic kinetic ($a=1/\sqrt{\pi n}$ is the 
average inter-particle distance). At zero temperature,
\begin{equation}
\label{eq.4}
E_{\rm kin}\simeq \hbar/m a^2,
\end{equation}
and $\tilde\Gamma_D = r_D = r_0/a, r_0 = m\mu^2/\hbar^2$ is the appropriate 
measure of the coupling strength; the characteristic length $r_0$ is the dipole 
equivalent of the Bohr radius. In fact, (\ref{eq.4}) is valid up to an arbitrary 
factor O(1): its exact evaluation would require information about the ground 
state momentum distribution. In the high-temperature classical domain, with
\begin{equation}
\label{eq.5}
E_{\rm kin} = 1/\beta,
\end{equation}
($\beta^{-1} = k_{\rm B}T$ is the thermal energy) and (\ref{eq.3}) becomes $
\Gamma_D = \beta\mu^2/a^3$, in analogy with the $\Gamma=\beta e^2/a$ coupling 
parameter for classical Coulomb liquids. The comparison of the two limits of $
\tilde\Gamma_D$ allows one to establish the correspondence between the high 
temperature classical) and low temperature (quantum) domains through the 
equivalence $\Gamma_D \Leftrightarrow r_D$.

The behavior of the dipole system can be analyzed over a wide range of 
temperatures and densities. The temperature domains may be identified through the 
value of the degeneracy parameter
\begin{equation}
\label{eq.6}
\Theta = \beta\hbar^2/2ma^2.
\end{equation}

As the quantum,$\Theta >1$ and classical, $\Theta <1$ and regimes. The 
equilibrium in the quantum domain exhibits a great richness of phenomena, such as 
condensation and quasi-condensation, superfluidity and crystallization, whose 
competition leads to an intricate low temperature phase diagram \cite{15}, quite 
at variance with the behavior in the classical domain. However, as far as the 
collective excitation spectrum is concerned, the situation is different: 
collective modes of many-body systems governed by the cooperative motion of many 
particles seem to be robustly insensitive to statistics (an obvious example is 
the plasmon in charged particle systems, which exhibits very much the same 
features in classical plasmas as in the degenerate electron gas).

We proceed now to compare the description of the collective mode dispersion based 
on two sets of computer simulation data. To the first set belong data pertaining 
to the static equilibrium properties of low temperature quantum dipole systems, 
through zero temperature quantum Monte Carlo (QMC) \cite{13} and low-temperature 
path integral Monte Carlo (PIMC) \cite{15} simulations. The second set of data 
originates from a Molecular Dynamics (MD) simulation of the dynamics of a 
classical point dipole system. The QMC simulation of the zero temperature quantum 
dipole system in the strong coupling regime has been pursued by Astrakharchik et 
al. \cite{13} and by Mazzanti et al. \cite{16}. In these works static structure 
factors have been generated as functions of the zero temperature coupling 
parameter $r_D$. The dispersion relation for the longitudinal dipole oscillations 
was obtained through two analytic approximations: first, through the Feynman 
formula (\ref{eq.1}), and, second, by incorporating the Feynman frequencies in a 
more sophisticated formalism based on the correlated basis function (CBF) 
approximation (extensively reviewed by Feenberg \cite{17}, Campbell \cite{18}, 
and Woo \cite{19}), which takes account of the effect of the three-phonon 
interaction term in the Hamiltonian. In the PIMC simulation approach \cite{20} 
followed by Filinov and co-workers \cite{15}, similar data for low but finite 
temperatures have been generated.	Coming now to the second set of data, a 
program of MD simulation of the dynamics of a classical point dipole system has 
been pursued by the authors, along the pattern used for simulating other strongly 
interacting classical systems \cite{21}. In this approach the dynamical structure 
function $S(k,\omega)$ is obtained for a range of values of the classical 
coupling parameter \cite{22} $\Gamma_D$ whose peaks reveal the collective 
excitation spectrum of the system. It should be emphasized that, in contrast to 
the work of Astrakarchik et. al. \cite{13}, this latter method provides direct 
information without the intervention of any further analytic approximation on the 
dispersion relation for the longitudinal dipole oscillations in the physical, 
albeit classical, system. Within the presently available computational technology 
it would be of course futile to expect data of the same kind to be created for 
quantum systems, where, to date, one has to rely on the intermediary of 
approximation techniques to connect the dynamics of the system with the available 
equilibrium information.

\begin{figure}
\onefigure[width=70mm]{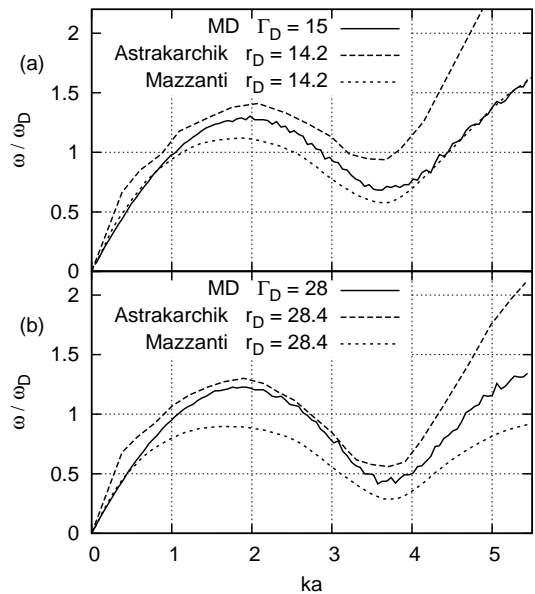}
\caption{Comparison of dispersion curves: MD (classical molecular dynamics), 
Astrakarchik \cite{13} (Feynman upper bound based on QMC), Mazzanti \cite{16} 
(presumed lower bound). (a) $\Gamma_D = 15$, (b) $\Gamma_D = 30$.}
\label{fig.1}
\end{figure}

We now compare the results from the two sets of data, which are presented in Fig. 
\ref{fig.1}, for approximately equivalent values of the classical and zero 
temperature coupling parameters $\tilde\Gamma_D$. The main statement of this 
paper derives from a number of conclusions that can be drawn from the inspection 
and analysis of these graphs.
\begin{enumerate}
\item There is no major difference in the morphologies of the dispersion curves 
pertaining to the quantum and classical systems; in particular, both of them show 
the formation of the roton minimum in the vicinity of $ka \simeq 3.5$.
\item Whether the roton minimum develops in the dispersion curve depends solely 
on the coupling parameter $\tilde\Gamma_D$: the minimum value for the curve to 
become non-monotonic is around $\tilde\Gamma_D = 5$, both for the classical and 
quantum system.
\item With increasing value of $\tilde\Gamma_D$, the roton minimum becomes 
deeper, reaching an absolute minimum. For the $T = 0$ curve this occurs at $
\tilde\Gamma_D = 28.6$, near the crystallization value \cite{13} $\tilde
\Gamma_D^*=30.2$.
\item The classical dispersion curve is sandwiched between the quantum Feynman 
and the quantum CBF curves; its being positioned below the Feynman curve is 
expected, since the latter constitutes an upper bound only. As to the CBF 
construction, it is difficult to assess its accuracy by analytic means; however, 
CBF calculations performed for $^4$He with model potentials have agreed well with 
measured spectra, with deviations attributable to the uncertainty of the actual 
interaction potential \cite{23}. Thus, one is probably on safe grounds by 
accepting the CBF curve as a lower bound.
\end{enumerate}

All these observations converge now towards a coherent physical picture that 
clearly suggests that the roton minimum is the consequence of strong 
correlations, which are basically a classical phenomenon, in the system. Quantum 
dynamics, Bose statistics, condensation, superfluidity, etc. seem to have very 
little influence either on the formation or on the structure of the roton 
minimum. It would be a strange coincidence indeed, if hybridization with vortex-
like or other unidentified excitations were responsible for the creation of the 
same phonon dispersion characteristics in bose liquids as they are manifested 
through strong correlations in the equivalent classical systems.

Having made our central observation, there are a few more issues to be addressed. 
Our discussion on the quantum system so far has been limited to the $T = 0$ 
ground state, whose analysis is sufficient to establish the correlational origin 
of the roton minimum. Finite temperatures are, however, now available: the recent 
PIMC calculations \cite{15} have generated $S(k)$ data in the $0.5<\Theta<3.3$ 
domain (on both sides of the superfluid/normal fluid transition point). The 
generalization of the zero temperature Feynman formula (\ref{eq.1}) to finite
temperatures becomes
\begin{equation}
\label{eq.7}
\omega(k)\tanh\left[\frac{\beta \hbar \omega(k)}{2}\right] = \frac{\hbar k^2}{2mS
(k)}.
\end{equation}

\begin{figure}
\onefigure[width=70mm]{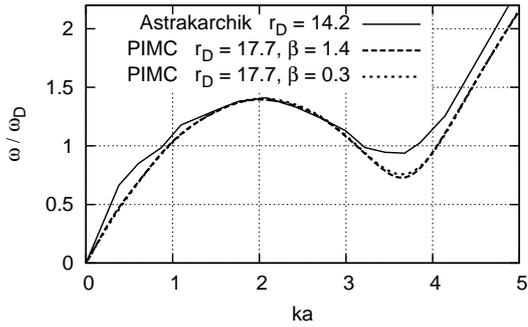}
\caption{Comparison of $T = 0$ and finite low temperature dispersions. Note that 
the finite temperature modifications are very small.}
\label{fig.2}
\end{figure}

The PIMC dispersion curves based on (\ref{eq.7}) for three representative $\Theta
$ values, together with the QMC ground state dispersion are displayed in Fig. 
\ref{fig.2}. The difference between the $T=0$ and $T\ne0$ dispersions is 
strikingly small (except near $k=0$, a region for which the inaccuracy of the QMC 
simulations \cite{13} has already been pointed out \cite{24}, further 
corroborating the expectation that as far as the dispersion properties of the 
collective excitations are concerned, the transition from the $T = 0$ degenerate 
to the $T \rightarrow \infty$ classical phase is smooth and eventless.

\begin{figure}
\onefigure[width=70mm]{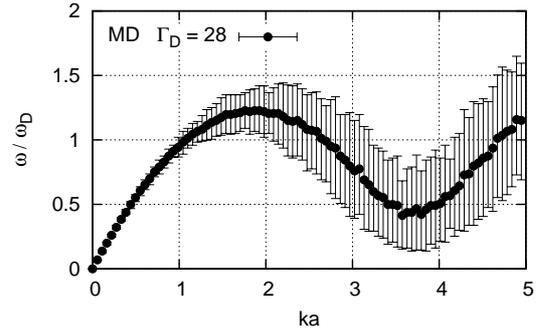}
\caption{Broadening of the phonon-maxon-roton line due to temperature effect. 
Error bars indicate the width (full width at half maximum) of the spectral peak.}
\label{fig.3}
\end{figure}

Quite at variance with the dispersion, one expects that the lifetime of the 
excitations (i.e. the damping of the oscillations) is strongly affected by rising 
temperature. Here one has to rely on what is known from observations on liquid 
$^4$He, which then can be compared with relevant information gleaned from the 
classical MD simulations. From the analysis of $S(k,\omega)$ with
the aid of neutron scattering experiments \cite{25} we know that the character of 
the roton minimum peak undergoes a dramatic change as the system passes from the 
superfluid to the normal phase: while it is very sharp in the superfluid phase, 
it becomes broad in the latter and as the temperature continues to rise above its 
critical value it becomes quite ill-defined. Thus one would expect a similarly 
broadened roton peak in the classical MD generated $S(k,\omega)$.
This indeed is the case. An examination of Fig. \ref{fig.3} that portrays $S(k,
\omega)$ and the width of the roton peak for a sequence of increasing $\Gamma_D$ 
values shows that the width decreases for higher coupling, yet remains 
considerable up to crystallization. Beyond observing the qualitatively identical 
behavior in the classical 2D dipole system and $^4$He, quantitative comparison 
would be quite meaningless, first because of the entirely different interaction 
potentials in the two systems, and, second, because studying the effect of the 
variation of the coupling strength is obviously not an available option in 4He 
experiments. The conclusion to be drawn from the comparison is, however, that the 
temperature engendered damping of the rotons is similar in the classical and 
quantum systems and its occurrence in no way invalidates the link between the 
finite $T$ classical and the $T \simeq 0$ quantum systems.

If we now accept the premise that the roton minimum experienced in the bose 
systems studied are caused by the strong correlations of classical origin, one 
may wonder about the generality of the link between strong correlations and 
roton-like behavior for other many- body systems. First, would systems governed 
by interaction potentials other than the ones that have been investigated exhibit 
a similar behavior? With a high level of confidence one can predict that this 
would indeed be the case. In fact, it was prior to the MD analysis of the dipole 
system that the classical "roton minimum" phenomenon was established for Coulomb 
and Yukawa systems \cite{21}. There is no reason to believe that the mapping of 
this behavior to the equivalent low temperature Bose system, should such a system 
become available for observation, would not occur in the same fashion as it does 
for a system of dipoles. Second, could the classical roton minimum behavior be 
mapped to a strongly interacting low temperature system of fermions in the same 
way as it is to a system of bosons? We do not have enough observational or 
theoretical understanding of such systems to provide a definite answer. 
Nevertheless, it is clear that the fundamentally different low temperature 
behaviors of the fermi and bose systems would make a similarity unlikely. The 
issue is that in a bose condensate or quasi-condensate the single particle 
excitations are suppressed, which makes the unimpeded development of the 
collective excitation, even in the high-$k$ domain where the roton minimum is to 
occur, possible. There is no such suppression in the fermi system. Taking the 
electron liquid as a paradigm, the high-$k$ development of the plasmon branch is 
profoundly affected by the presence of the electron-hole pair continuum. Based on 
a simplified model \cite{26}, with a measure of caveat it can be argued that the 
pair (and multi-pair) continuum not only generates a strong damping to the 
collective mode, but also inhibits its evolution in the high-$k$ domain. Thus it 
seems safe to conclude that the condensed (or quasi- condensed) bose system is 
the sole candidate for the clear manifestation of the correlation induced roton 
minimum.

The final issue we want to address is what has become to be known in the $^4$He 
studies as the issue of the two-roton state. It was shown first by Cowley and 
Woods \cite{27} that in addition to the phonon excitation, a high frequency 
branch can be identified in the observed mode spectrum of the liquid $^4$He: this 
mode was identified as a two roton excitation, with a frequency in the vicinity 
of twice the frequency of the roton minimum and rather insensitive to the wave 
number. More recent neutron scattering studies \cite{28} have found additional 
branches with higher frequencies that can be attributed to roton-maxon and maxon-
maxon excitations.

The identification of these high frequency states in terms of the excitation of 
two quasi-particles goes back to the classic paper of Miller, Pines and Nozieres 
\cite{29}. In the original model, the quasi-particles were non-interacting: the 
effect of the interaction between them was suggested by Pitaevski \cite{30} and 
first considered in detail in the pioneering series of works by Ruvalds, 
Zawadowski, Bedell , Pines and collaborators \cite{30}. In these works, depending 
on the sign of the interaction (attractive vs. repulsive), a downward or upward 
shift from the simple additive nominal values of the high frequency excitations 
is predicted.

The generation of higher harmonics and combination frequencies is a well-known 
nonlinear effect \cite{31}, due to the inherent nonlinearity of the interaction 
between the constituent ($^4$He, dipoles e.g.) particles. Given a dispersion 
curve, any pair of frequencies along the curve is a candidate for the generation 
of combination frequencies: it is, however, the frequencies in the vicinity of 
extrema of the dispersion curve, where the density of states is the highest, 
which are primarily responsible for creating observable new branches of 
collective excitations. In this paper we have argued that the generation of the 
roton minimum and of the maxon maximum in strongly coupled bose condensate 
(quasi-condensate) liquids is a universal effect due the strong interaction 
between their constituents. A corollary to this statement is now that the roton-
roton, roton-maxon and maxon-maxon states should also be identifiable for the 
equivalent classical liquid through the appearance of a significant weight in the 
density correlation spectra at the corresponding combination frequencies.

\begin{figure}
\onefigure[width=80mm]{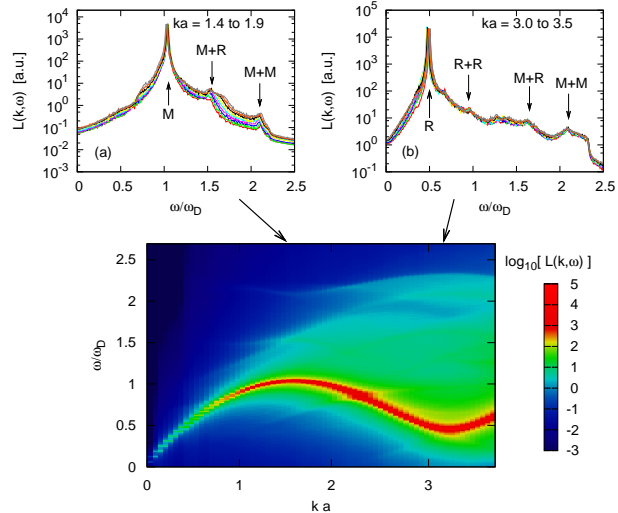}
\caption{(color online) Illustration of the appearance of the R+R, R+M, M+M 
combination frequencies in the longitudinal current-current correlation functions 
$L(k,\omega)$, measured along the direction of the nearest neighbor in a 
hexagonal lattice). The center color map of the $L(k,\omega)$ fluctuation spectra 
shows the strong primary dispersion and the "ghosts" of the combination 
frequencies. The small panels show vertical cross-sections taken for the selected 
group of wave-numbers, where the appearance of the combination frequencies is the 
most manifest.}
\label{fig.4}
\end{figure}

The MD analysis of $S(k,\omega)$ of the strongly coupled dipole liquid when 
extended into the high frequency regime up to and beyond twice the maxon 
frequency indeed reveals these expected features. In Fig. \ref{fig.4} we have 
displayed a series of $S(k,\omega)$ spectra for a wide range of k values. In 
order to emulate the low level of the background noise characteristic of the 
superfluid, the
graphs were taken at $\Gamma_D=500$, well beyond the crystallization value, but 
at a value of coupling where the thermal motion is of sufficient amplitude to 
generate nonlinear effects. Thus these graphs may be considered appropriate to 
illustrate a trend, but a detailed analysis of the features exhibited in them is 
not our purpose here: it is sufficient to point out that the essential feature, 
namely the accumulation of weight in the vicinity of three combination 
frequencies, the roton-roton (R+R), roton-maxon (R+M),and maxon-maxon (M+M) is 
clearly visible. From other works we know that the generation of harmonics, 
similarly to the development of the roton minimum itself, appears to be a general 
feature of strongly coupled liquids with Yukawa, Coulomb and other similar types 
of interaction \cite{32} as well. The relative amplitudes of these harmonics are 
very much functions of the particular type of interaction potential prevailing in 
the system. Thus, attempts at a quantitative comparison with experimental data on 
liquid $^4$He would in all likelihood not be useful. On the other hand, the 
positions of the combination frequencies, i. e. their shifts from their nominal 
values, is due, in the language of quasiparticle states, to the interaction 
between the quasiparticles and, as such, their comparison with the predictions of 
the RZ theory as applied to the 2DDS would certainly be of interest.

To summarize, in this paper we have provided compelling support for two earlier 
suggestions \cite{7,8}, that the formation of maxon-roton excitations in 
condensed or quasi-condensed bose systems is the result of a basically classical 
strong correlational effect. Our assessment is based on the analysis of the 
collective mode dispersion of the strongly coupled 2D bosonic dipole system 
(2DDS) below and above the superfluid transition temperature \cite{13,14}. This 
dispersion exhibits a phonon-maxon-roton structure, very much the same as the 
collective mode dispersion in $^4$He. Paralleling these studies, we investigated 
\cite{22} the dispersion of the strongly coupled 2DDS in the high-temperature 
classical domain through molecular dynamics (MD) simulations: we have discovered 
a remarkable congruence between the collective mode dispersions of the classical 
and quantum systems, pointing at a common origin of these phenomena, the shaping 
of the longitudinal phonon dispersion by strong interparticle correlations. 
Models that seek the origin of the roton in excitations different in nature from 
the phonon seem to be incompatible with this evidence.	Moreover, the MD 
simulations now can be trusted to provide a first-time glimpse into the actual 
behavior of the dynamics of the strongly correlated system without reliance on 
the intermediary of models. The approach that describes correlational effects in 
the dynamical behavior through the static structure function is superseded by the 
direct dynamical MD simulations. Furthermore, our observations of a considerable 
weight in the MD density fluctuation spectrum in the vicinity of the roton-roton, 
roton-maxon, and maxon-maxon combination frequencies, similar to what has been 
observed in neutron scattering experiments \cite{28} in $^4$He, may be regarded 
as additional corroboration of the underlying strong correlational classical 
model.

%\begin{figure}
%\onefigure[width=80mm]{figure1.eps}
%\caption{}
%\label{fig.1}
%\end{figure}

\acknowledgments

We wish to acknowledge useful discussions with A. Zawadowski. This work has been 
partially supported by OTKA-PD-75113, OTKA- K-77653, MTA-NSF/102, NSF Grants 
PHY-0813153, PHY-0812956 and PHY-0903808; it was also supported by the Janos 
Bolyai Research Grant of the Hungarian Academy of Sciences.

\end{document}